\documentclass[12pt]{article}
\usepackage{amsmath, amssymb, breqn}
\usepackage[utf8]{inputenc}
\usepackage{caption}
\usepackage{subcaption}
\usepackage{float}
\usepackage{array}
\usepackage[LGR,T1]{fontenc}
\usepackage{eurosym}
\usepackage{fancyhdr}
\usepackage[countmax]{subfloat}
\usepackage{booktabs}
\usepackage{subfig}
\usepackage{appendix}
\usepackage{amsmath}
\usepackage{amssymb}
\usepackage{graphicx}
\usepackage{graphics}
\usepackage{nicefrac}
\usepackage{xcolor}
\usepackage[authoryear]{natbib}
\begin{document}

\begin{center}
\textbf{Symbiosis emergence and abandonment in nature: a coordination game
approach\footnote{Correspondence to [tedt@rice.edu]}}

\vspace{1cm}

Simon A. Levin\footnote{Princeton University}

\bigskip

Ted Loch-Temzelides\footnote{Rice University}

\bigskip

2025

\bigskip

\textbf{Abstract}
\end{center}

\noindent \noindent We employ an \textit{n}-player coordination game to
model mutualism emergence and abandonment. We illustrate our findings in the
context of the host--host interactions among plants in plant-mycorrhizal
fungi (MF) mutualisms. The coordination game payoff structure captures the
insight that mutualistic strategies lead to robust advantages only after
such \textquotedblleft biological markets" reach a certain scale. The game
gives rise to three types of Nash equilibria, which correspond to the states
derived in studies of the ancestral reconstruction of the mycorrhizal
symbiosis in seed plants. We show that all types of Nash equilibria
correspond to steady states of a dynamical system describing the underlying
evolutionary process. We then employ methods from large deviation theory on
discrete-time Markov processes to study stochastic evolutionary dynamics. We
provide a sharp analytical characterization of the stochastic steady states
and of the transition dynamics across Nash equilibria and employ simulations
to illustrate these results in special cases. We find that the mutualism is
abandoned and re-established several times through evolutionary time, but
the mutualism may persist the majority of time. Changes that reduce the
benefit-to-cost ratio associated with the symbiosis increase the likelihood
of its abandonment. While the mutualism establishment and abandonment could
result from direct transitions across the mutualistic and non-mutualistic
states, it is far more likely for such transitions to occur indirectly
through intermediate partially mutualistic states. The MF-plant mutualism
might be (partially or fully) abandoned by plants even if it provides
overall superior fitness.

\newpage

\section{Introduction}

Though mutualisms are common in nature, there is no general theory on
whether they will be stable once they have evolved and on what drives the
observed repeated transitions between mutualistic and non-mutualistic
states. For example, pollination mutualisms in angiosperms appear to be lost
frequently (Culley et al. 2002; Friedman 2011), whereas the mutualistic
symbiosis between plants and nitrogen-fixing bacteria appears to persist
over evolutionary time (Werner et al. 2014). Other examples of mutualisms
that are formed, dissolved, and possibly re-established over time include
the following.

\textbf{Ant--Plant Mutualisms}

The symbiosis between ants and plants involves many species throughout the
tropics and was one of the first mutualisms to be investigated by
ecologists. Plant--insect mutualisms have arisen and been lost repeatedly
(Bronstein et al, 2006). Many Acacia species have mutualistic relationships
with ants (e.g., Pseudomyrmex), providing shelter and food to several
species of ants who in turn defend the trees from herbivores. Palmer et al.
(2008) investigated the effects of large mammalian herbivores on an
ant-Acacia mutualism in an African savanna. In the absence of browsing by
large herbivores, A. drepanolobium trees stopped producing nectar and hollow
thorns, essentially abandoning the mutualism when ant protection became less
beneficial. Such mutualisms can re-evolve if herbivore threats return or as
host trees develop and abandon repeated symbiotic relationships with
different competing species of ants (Palmer et al. 2013).

\textbf{Coral--Algae (Zooxanthellae) Mutualisms}

Scleractinian (stony) corals form symbioses with a wide range of symbiotic
algae, including phototrophic dinoflagellates in the genus Symbiodinium, for
nutrients via photosynthesis. Under thermal stress, bleaching might offer an
opportunity for reef corals to rid themselves of suboptimal alga,
temporarily abandoning the symbiosis. As the coral host depends on
photosynthate for nutrition, a prolonged breakdown of the symbiosis can lead
to coral death (Baker, 2003). However, corals may re-establish the
mutualism, sometimes with more heat-tolerant algae variants (Baker, 2001).
Thus, switching to more thermally tolerant symbionts has the potential to
benefit coral reefs that face increasingly frequent mass bleaching due to
climate change. Boulotte el al (2016) found evidence for symbiont switching
in reef-building corals, with two de novo acquired thermally resistant
Symbiodinium types, suggesting that this switching may have been driven by
consecutive thermal bleaching events. While these changes involve relatively
short horizons, they do correspond to transitions across different
equilibrium outcomes and can be modeled using coordination games.

\textbf{Land plant-arbuscular mycorrhizal fungi mutualisms}

Land plant-arbuscular mycorrhizal fungi mutualisms are sometimes abandoned,
partial, re-established, etc. over evolutionary time (Werner et al. 2018).
Figure 1 represents derived evolutionary transitions rates across
plant-mycorrhizal fungi mutualistic states (\textbf{AM}), non-mutualistic
states (\textbf{NM}), and partially mutualistic states (\textbf{AMNM}) for
monocots (Maherali et al, 2016). The number in parentheses indicates the
percent of species in that state, while the number next to each arrow
indicates the transition rate in numbers of transitions per million years.
In what follows, we will attempt to understand the transitions between
mutualistic, partially mutualistic, and non-mutualistic states as the
outcome of an evolutionary process operating on an underlying \emph{%
coordination game}. The numbers in parenthesis will then correspond to the
fraction of time spent by plants in each respective state over a long time
horizon.

\begin{figure}[!htpb]
\includegraphics[width=1.0\textwidth]{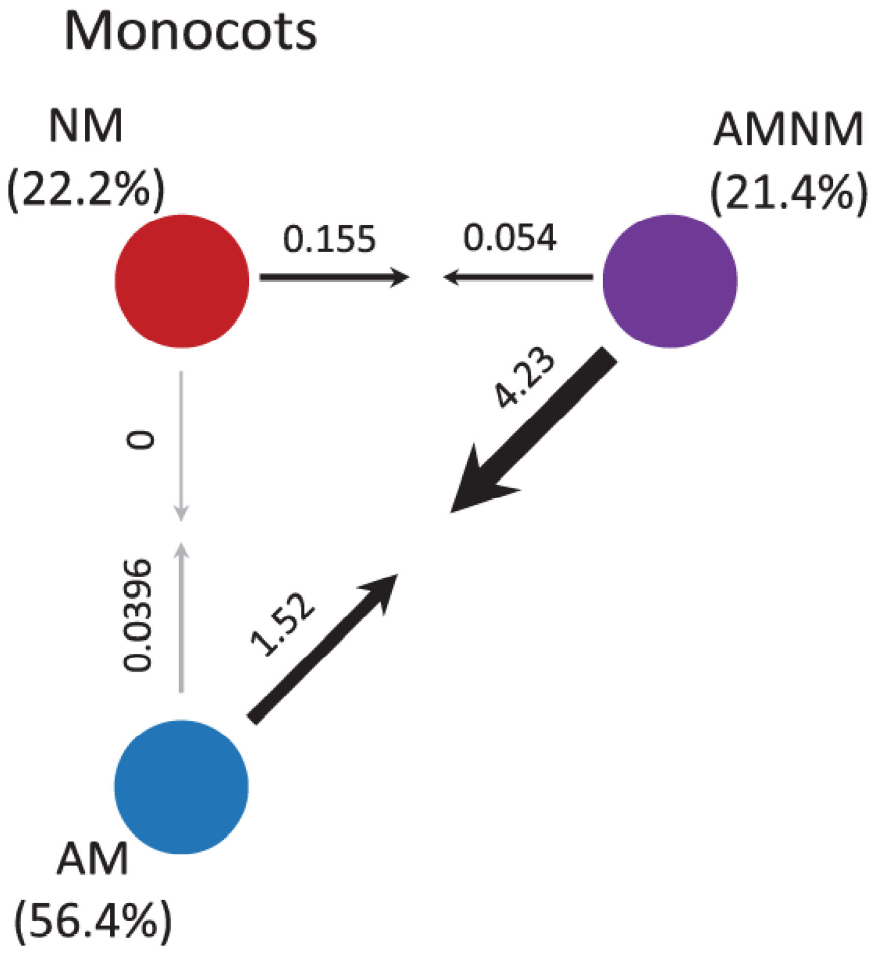}
 \centering
\caption{Transitions between mutualistic (AM), partially mutualistic (AMNM) and non-mutualistic (NM) states (Figure from Maherali et al, 2016)}
\label{fig:fig1}
\end{figure}

Our \emph{coordination game} captures the qualitative features of the relative
payoffs associated with the symbiosis and its abandonment. We will then
employ methods from large deviation theory to characterize analytically the
evolutionary dynamics and the associated long-run outcomes of an
evolutionary game. As an illustration, we will frame the model in the
context of the plant-mycorrhizal fungi (MF) mutualisms (Maherali et al.
2016). More precisely, we model the host plants as playing a coordination
game in evolutionary time. As in Halloway et al (2022), we will take the
behavior of their MF as fixed and in what follows we will concentrate on the
host interactions between the seed plants.

Previous studies of mutualism persistence and abandonment using evolutionary
game theory have concentrated on the role of cheating; see, for example,
Bronstein (2001), Ferriere et al (2002), Bronstein (2006), and Jones et al
(2015). Here we provide an alternative game-theoretic explanation where
mutualism abandonment can take place even in the absence of cheating, for
example, in cases where effective punishment strategies have eliminated any
advantage to cheaters. Coordination games offer an interesting paradigm, as
they give rise to multiple Pareto ranked Nash equilibria\textit{\ }(Nash,
1951).\footnote{%
Two equilibria are Pareto ranked if one yields strictly higher payoffs for
all players than the other.} They have been used in economics and the social
sciences to model market creation, new technology adoption, economic
recessions, and social conventions, among other phenomena. In the context of
biological mutualistic markets (No\"{e} and Hammerstein, 1994, 1995) the
coordination game payoff structure captures the insight that such markets
cannot operate effectively, unless they reach a certain participation size
or \textquotedblleft \textit{thickness}." That is, establishing or
reintroducing a mutualism might not be viable unless it reaches a certain
scale in terms of the number of individuals participating. Once that
threshold is reached, mutualistic strategies can lead to robust advantages.
This can be, for example, due to diversification and the ability to adjust
to changes in external conditions, both of which require the biological
market to operate at a certain minimal scale.

The model gives rise to three types of Nash equilibria, which match the
states derived in the formal ancestral reconstruction of the mycorrhizal
symbiosis in seed plants in Maherali et al (2016). The first corresponds to
the state of full plant-mycorrhizal fungi mutualism (\textbf{AM}). The
second corresponds to a complete mutualism abandonment (\textbf{NM}).
Finally, the third type captures mixed equilibria of partial mutualisms (%
\textbf{AMNM}). In terms of dynamics, all three types of Nash equilibria
correspond to steady states of the deterministic dynamical system describing
the underlying evolutionary process. Introducing random perturbations
(\textquotedblleft mutations") allows us to study the resulting stochastic
evolutionary dynamics. Employing techniques from large deviation theory in
the study of discrete-time Markov processes; see, for example, Freidlin and
Wentzell (1984), Kandori et al (KMR, 1993), Young (1993), Ellison (2000),
Blume et al (2003), and Dembo and Zeitouni (2009), we provide a sharp
analytical characterization of the stochastic steady states (ergodic
distributions) and of the transition dynamics across Nash equilibria in
evolutionary time. We use simulations to illustrate these results in special
cases.

Our main finding is that the evolutionary coordination game qualitatively
captures the main features of the time evolution of the plant-MF symbiosis
discussed in Maherali et al (2016). Notably, our model is consistent with
the following observations: \textit{(i)} the mutualism is abandoned and
re-established several times through evolutionary time, \textit{(ii)} the
mutualism persists the majority of time in the seed plant-mycorrhizal
symbiosis, (\textit{iii}) environmental and other changes that lower the
benefit-to-cost ratio of a symbiosis increase the likelihood of its
abandonment and can therefore serve as indicators of such transitions, and (%
\textit{iv}) while the symbiosis establishment and abandonment could result
from direct transitions across the \textbf{AM} and the \textbf{NM} states,
it is far more likely for such transitions to occur indirectly through
intermediate \textbf{AMNM} states. Interestingly, the MF-plant mutualism
might be (partially or fully) abandoned even if it is overall superior for
plant fitness. Concepts from the theory of large deviations tailored to
discrete-time Markov chains, notably the \textit{modified co-radius}
(Ellison, 2000), formalize the notion that large evolutionary changes driven
by random mutations are more likely when they can be achieved by passing
through a number of \textquotedblleft transient" steady states. We
characterize conditions that guarantee that over a long enough time horizon,
the system spends most of the time in the \textbf{AM}\ Nash equilibrium.
However, the system will eventually escape and reach another limit set,
before it escapes again over evolutionary time. The resulting dynamics also
explains why the \textbf{AM} state is reached more frequently through the
intermediary \textbf{AMNM} state.

Archetti et al (2011) discuss applications of economic game theory,
including signalling, principal-agent models, and models involving public
goods, to the study of mutualisms. They do not, however, discuss stochastic
evolutionary dynamics in the context of coordination games, which is the
focus of our study. McNickle and Dybzinski (2013) provide an accessible
general introduction to some of the concepts of game theory for plant
ecologists. Xu et al (2021, 2023) employ landscape-flux methods from
non-equilibrium statistical mechanics to investigate transitions across
locally stable steady states in an ecological context. Our approach differs
from theirs in two main ways. First, the dynamics in our model is driven by
an underlying coordination game whose Nash equilibria are in an one-to-one
correspondence with the steady states of the dynamical system employed to
capture the evolution of a mutualism. Second, although the model gives rise
to multiple steady states, we employ large deviation Markov chain methods to
derive a sharp characterization of the long-run behavior of the system in
the presence of infrequent random perturbations, or \textquotedblleft
mutations."

Of special interest for our analysis is the study by Halloway et al (2022).
They employ a game-theoretic framework to study the symbiosis between plants
and their microbial symbiotes. Like us, they focus on host--host
interactions among host plants. Their analysis is based on $2$-player games,
where they investigate the possibility of coexistence of mutualist and
non-mutualist strategies in the plant population. The payoff matrix in their
game captures the insight that a larger fraction of symbiotic host plants
can make the microbial symbiosis \textit{less} beneficial if resources are
limited, thus reducing the usefulness and frequency of mutualism. Their
model predicts that mutualist and non-mutualists frequently coexist at the
same time within a population. The coordination game structure, in our $n$%
-player coordination game attempts to capture a different force, namely that
once a certain participation threshold needed to establish the biological
market is reached, the benefits of a mutualism \textit{increase} in the
number of participants. Like in other markets, this would be true if, for
example, increased market thickness results in more reliable supply and
resilience to outside shocks. Employing techniques from stochastic
evolutionary dynamics allows for multiple \textit{over time transitions}
across the \textbf{AM}, \textbf{NM}, and \textbf{AMNM} Nash states in our
model.

In what follows, after discussing the coordination game structure, we will
introduce deterministic, followed by stochastic evolutionary dynamics. The
latter will result in repeated transitions across the static Nash equilibria
over evolutionary time, mimicking the historical record of the repeated
abandonment and reintroduction of the biological market mutualism. We then
use simulations to illustrate these results in special cases. The SI
Appendix contains a formal treatment of the stochastic evolutionary dynamics.

\section{The \textit{n}-player\textit{\ }coordination game and Nash
equilibrium}

To understand the processes through which the mycorrhizal symbiosis is
maintained or lost, Maherali et al (2016) reconstructed its evolution using
an approximately 3,000-species seed plant phylogeny integrated with
mycorrhizal state information. For our purposes, their analysis identifies
the following qualitative features: \textit{(i)} \textbf{AM} symbiosis is
persistent; \textit{(ii)} direct transitions between \textbf{AM} and \textbf{%
NM} states are rare, indicating that evolutionary forces favor stasis when
one of these states is reached, and that mutations that allow transitions
between states occur at a relatively low rate; \textit{(iii)} reversions
from \textbf{AMNM} back to \textbf{AM} are an order of magnitude more likely
than transitions to the \textbf{NM} state, suggesting that natural selection
favors \textbf{AM} symbiosis over mutualism abandonment; and \textit{(iv)}
the transition rates from \textbf{NM} to \textbf{AMNM} are higher than the
reverse, thus, loss of mycorrhizal symbiosis can be recovered through the
mixed \textbf{AMNM }states. We will demonstrate that the coordination
game-theoretic framework we will introduce provides a mechanism that
captures several of these features.

We consider the plant-MF mutualism as an example of a \textquotedblleft 
\textit{biological market"} (see No\"{e} and Hammerstein, 1994, 1995), in
which plants supply carbon to MF in exchange for nutrients. Like any other
market, biological markets require a certain level of \textquotedblleft
thickness" to be viable. In other words, if participation is not
sufficiently large, the market is unlikely to be a reliable source of the
desirable commodities, as small changes, for example in underlying
environmental conditions, could lead the mutualism to collapse in favor of
alternative ways to obtain the necessary commodities. This feature is
captured by the notion of a \textit{coordination game. }In what follows, we
will employ a \textquotedblleft partial equilibrium" approach. We will take
the behavior of the fungi as exogenously given, and will concentrate on the
game played by a population of plants. Like the prisoner's dilemma game, the
coordination game paradigm is a well-studied model in economics, but not
often studied in biology. Unlike the prisoner's dilemma game, which obtains
a unique Nash equilibrium in dominant strategies, a coordination game gives
rise to multiple Nash equilibria. These equilibria have a natural
correspondence to the observed outcomes of the biological market mutualism.
The first Nash equilibrium corresponds to the outcome where the symbiosis is
established (\textbf{AM}). The second Nash equilibrium corresponds to the
outcome where the symbiosis is abandoned (\textbf{NM}). It is worth pointing
out that both pure-strategy Nash equilibria are \textit{strict}, therefore
they constitute evolutionary stable strategies (ESS).\footnote{%
A Nash equilibrium is strict if each player's strategy is their unique best
response, meaning any deviation would make them strictly worse off.} In the
case of the \textbf{AM} equilibrium, a mutant that abandons the
well-established symbiosis would be worse off. Similarly, in the \textbf{NM}
equilibrium, where the symbiosis is non-existent, a mutant (or indeed a
small number of mutants) would not be able to create the market thickness
necessary for the symbiotic biological market to take off. Finally, the
mixed Nash equilibria (\textbf{AMNM}) correspond to cases where the
symbiosis is pursued by a sufficient number of plants to create a (barely)
functional biological market, which is as good as other alternatives. In
this case, only a fraction of the plant population pursues the symbiosis. We
remark that the \textbf{NM} Nash equilibrium is an ESS even though the 
\textbf{AM}\ equilibrium is associated with a higher plant population
fitness. Thus, coordination games can explain why evolution might lead to
mutualism abandonment and stable outcomes of inferior overall fitness even
in the absence of \textquotedblleft cheating."

Formally, we consider a symmetric normal-form game, $\Gamma =\langle
N,S^{i},u^{i}\rangle $, with $n$ identical players and two strategies. The
set $N=\{1,2,\ldots ,n\}$ denotes the set of players. The $n$ players
correspond to seed plants of a given type. Let $S=\left\{
s_{1},s_{2}\right\} $ be the set of pure strategies; $s_{1}$ stands for
\textquotedblleft engage in plant-mycorrhizal fungi mutualism (\textbf{AM}%
)," while $s_{2}$ stands for \textquotedblleft do not engage in
plant-mycorrhizal fungi mutualism (\textbf{NM})." The payoff functions for
each player are identical, meaning $u^{i}=u$ for all $i\in N$. The payoff
function $u$ is defined as follows. Let $\mathbf{s}=(s^{1},s^{2},\ldots
,s^{n})$ be a \textit{strategy profile}; i.e., a vector describing the
strategies played by each player $i$, where $s^{i}\in \{s_{1},s_{2}\}$, for
each $i$. We use $\mathbf{s}^{-i}=(s^{1},...,s^{i-1},s^{i+1}\ldots ,s^{n})$
to denote the strategy profile of everyone but player $i$. We have the
following.

\noindent \textbf{Definition 1:} \textit{A strategy profile }$\widehat{%
\mathbf{s}}$\textit{\ is a pure strategy Nash equilibrium for }$\Gamma $%
\textit{\ if, for all }$i$\textit{, }$u^{i}(\widehat{\mathbf{s}})\geq
u^{i}(s^{i},\widehat{\mathbf{s}}^{-i})$\textit{, for all }$s^{i}\in S^{i}$%
\textit{. A Nash equilibrium }$\widehat{\mathbf{s}}$\textit{\ is strict iff
for all }$i$\textit{, }$u^{i}(\widehat{\mathbf{s}})>u^{i}(s^{i},\widehat{%
\mathbf{s}}^{-i})$\textit{, for all }$s^{i}\in S^{i}$\textit{.}

Let $z$ be the number of players employing strategy $s_{1}$; thus $n-z$ is
the number of players employing strategy $s_{2}$. The payoff for each player
depends on the number of players employing each strategy. Let $u(s_{i},z)$
denote the payoff for a player employing strategy $s_{i}$ when $z$ other
players also employ $s_{1}$. Formally, we have the following.

\noindent \textbf{Definition 2:} \textit{The game} $\Gamma $ \textit{is a
coordination game if the following conditions hold:}

\textit{(C1) Highest payoffs for coordination in strategy}\textbf{\ }$s_{1}$%
\textit{:} 
\begin{equation}
u(s_{1},n)>u(s_{2},n-k)\text{,}\quad \text{for all}\quad 0\leq k\leq n
\end{equation}

\textit{(C2) Increasing payoffs with coordination in either strategy:} 
\begin{equation}
u(s_{1},z+1)\geq u(s_{1},z)\quad \text{and}\quad u(s_{2},z)\geq u(s_{2},z+1)%
\text{, for all }0\leq z<n-1
\end{equation}

\textit{(C3) Coordination threshold region:}\textbf{\ }

\noindent $\exists z_{\ast }$, $z^{\ast }$ \textit{such that}

\begin{equation}
u(s_{1},z)>u(s_{2},z)\text{,}\quad \text{\textit{for all}}\quad 0<z^{\ast
}<z\leq n-1
\end{equation}

\begin{equation}
u(s_{1},z)=u(s_{2},z)\text{,}\quad \text{\textit{for all}}\quad 0<z_{\ast
}\leq z\leq z^{\ast }<n-1
\end{equation}

\begin{equation}
u(s_{1},z)<u(s_{2},z)\text{,}\quad \text{\textit{for all}}\quad 0\leq
z<z_{\ast }<n-1
\end{equation}%

\begin{figure}[!htpb]
\includegraphics[width=1.0\textwidth]{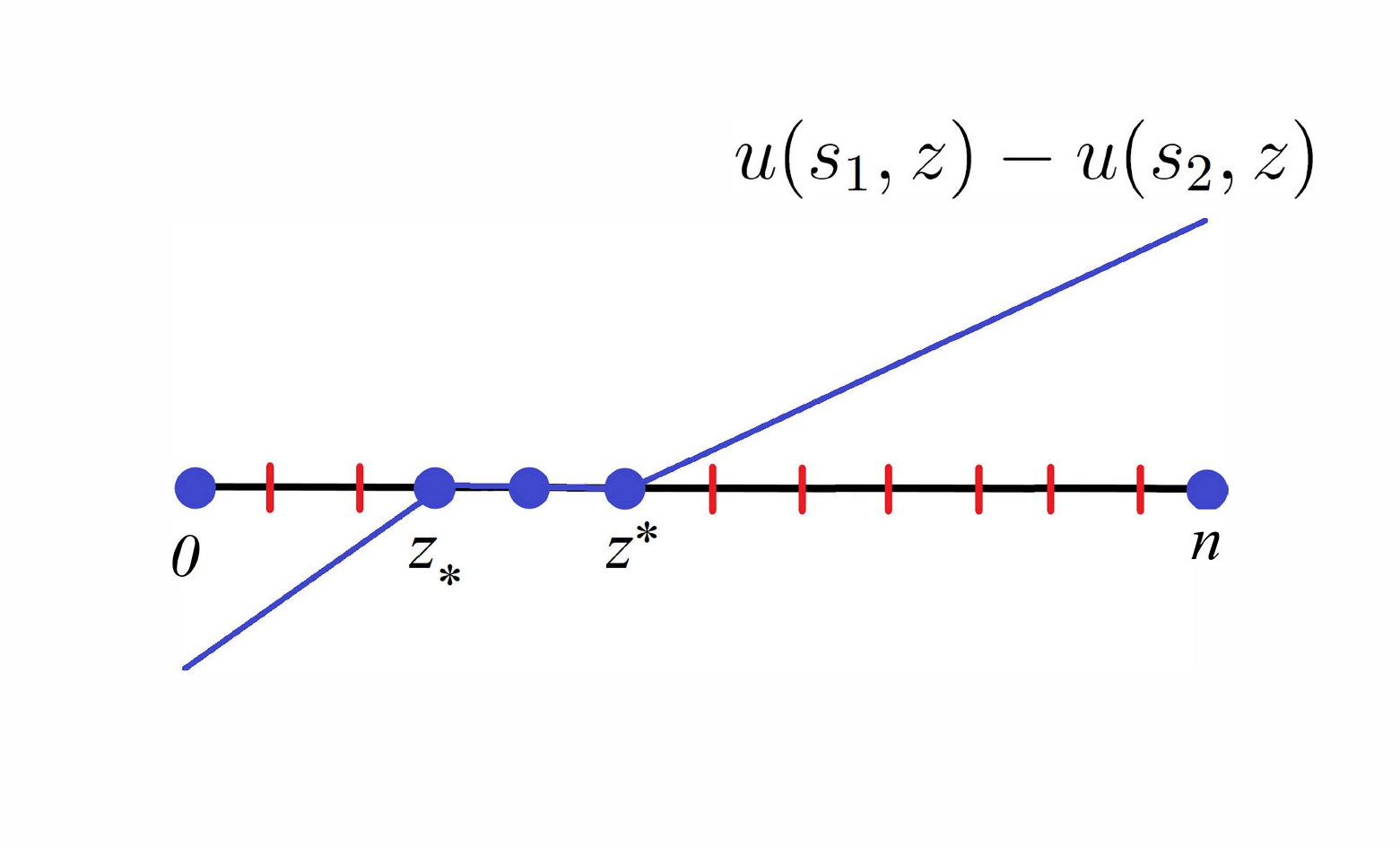}
 \centering
\caption{Coordination game: payoffs and basins of attraction}
\label{fig:fig2}
\end{figure}

Condition (C1) implies that strategy profile $\widehat{\mathbf{s}}%
_{1}=(s_{1},...,s_{1})$ is the Pareto optimal pure strategy Nash equilibrium
of the game. Condition (C2) implies that the payoff from choosing strategy $%
s_{1}$ (respectively $s_{2}$) weakly increases as more players choose
strategy $s_{1}$ (respectively $s_{2}$). Condition (C3) implies the
existence of a set of mixed Nash equilibria indexed by $z\in \lbrack z_{\ast
},z^{\ast }]$. The two pure strategy Nash equilibria of the game $\widehat{%
\mathbf{s}}_{1}=(s_{1},...,s_{1})$ and $\widehat{\mathbf{s}}%
_{2}=(s_{2},...,s_{2})$ are strict (therefore, ESS). The thresholds $z_{\ast
}$, $z^{\ast }$ will play an important role when we discuss dynamics, as
they define the basins of attraction of the respective Nash equilibria. More
concretely, when $z>z^{\ast }$ enough plants coordinate on the symbiotic (%
\textbf{AM}) strategy to make it the one providing higher fitness. In
contrast, when $z<z_{\ast }$ the symbiosis is not viable, as it is not
pursued by enough plants, making symbiosis abandonment (\textbf{NM}) the
more successful strategy. Finally, $z\in \lbrack z_{\ast },z^{\ast }]$
represents the mixed set of states (\textbf{AMNM}) where the symbiosis is
pursued by just enough plants for the relative payoffs of the two strategies
to be effectively equal.

Figure 2 gives an illustrative example of the difference in payoffs, $%
u(s_{1},z)-u(s_{2},z)$, that is consistent with the coordination game
assumptions. The payoff difference between strategy $s_{1}$and strategy $%
s_{2}$ is positive for high values of $z$, and negative for low values of $z$%
. This results in the two pure-strategy Nash equilibria at $z=0$ and at $z=n$%
. Note that for $z\in \lbrack z_{\ast },z^{\ast }]$ we have $%
u(s_{1},z)-u(s_{2},z)=0$, thus, states $z\in \lbrack z_{\ast },z^{\ast }]$
correspond to mixed Nash equilibria. We shall return on this example later,
after we introduce dynamics.

\section{Evolutionary dynamics}

We now turn to the relationship between Nash equilibria and the steady
states of a dynamical system describing the underlying evolutionary process
taking place in discrete time, $t=0,1,2,...$. The state, $z_{t}$, gives the
number of players adopting strategy $s_{1}$ at time $t$; $z\in
Z=\{0,1,...,n\}$. The state evolution is described by a dynamical system:

\begin{equation}
z_{t+1}=b(z_{t})
\end{equation}%
The function $z_{t+1}=b(z_{t})$ gives the strategy representation in the
population at $t+1$, given that the time $t$ state is $z_{t}$. We assume the
following weak monotonicity \textquotedblleft Darwinian" property and
boundary conditions (see, for example, KMR, 1993):

\begin{eqnarray}
sign\left( b(z)-z)\right) &=&sign[u(s_{1},z)-u(s_{2},z)]\text{, if }0<z<n 
\notag \\
b(0) &=&0,\text{ }b(n)=n
\end{eqnarray}%
In words, the Darwinian property only requires that the more successful
strategy increases its representation in the population in the next period.
Examples of such dynamics include the \textit{(myopic)} \textit{best reply
dynamic}, often used in economics:$\qquad \qquad \qquad \qquad \qquad $%
\begin{equation}
BR(z)=\left\{ 
\begin{array}{ccc}
n, & \text{if} & u(s_{1},z)>u(s_{2},z) \\ 
z, & \text{if} & u(s_{1},z)=u(s_{2},z) \\ 
0, & \text{if} & u(s_{1},z)<u(s_{2},z)%
\end{array}%
\right.
\end{equation}%
and the \textit{replicator dynamic} used in biology:\footnote{%
The familiar continuous-time version of the replicator dynamic would read: $%
\overset{.}{z}=z[u(s_{1},z)-\overline{u}(z)]$, where $\overline{u}(z)$ is
the average payoff over all strategies given population configuration $z$.
In other words, a strategy that results in higher than average fitness
increases its representation in the population over time, and vice versa;
see, for example, Taylor and Jonker (1978), Maynard Smith (1982), Hofbauer
and Sigmund (1998).}

\begin{equation}
\mathrm{B}(z)=z\left[ \frac{u(s_{1},z)}{zu(s_{1},z)+(n-z)u(s_{2},z)}\right]
\label{R}
\end{equation}%
In the context of a coordination game, any deterministic dynamical system
satisfying the above monotonicity property has two pure steady states, $0$
and $n$, as well as mixed steady states \textit{\ }$z\in \lbrack z_{\ast
},z^{\ast }]$, corresponding exactly to the pure (\textbf{AM},\textbf{NM})
and the mixed (\textbf{AMNM}) Nash equilibria of stage game, respectively.
The long-run behavior of the deterministic dynamic depends on the initial
distribution across states. Initial states $z_{0}$ such that $z_{0}>z^{\ast
} $ will converge to the \textbf{AM} steady state, while initial states $%
z_{0}$ such that $z_{0}<z_{\ast }$ will converge to the \textbf{NM} steady
state (see Figure 2). Furthermore, a deterministic system does not permit
transitions across steady states, which are the focus of our analysis.

The dependence on initial conditions is resolved if noise or
\textquotedblleft mutations" are introduced into the system. We assume that
with probability $\epsilon $, each player mutates, playing each strategy
with probability $1/2$. Mutations are assumed to be $iid$ across players and
time. This yields a stochastic dynamical system on the finite state space $Z$%
. The associated stochastic evolutionary dynamics gives rise to a Markov
chain with a unique invariant distribution, $\mu (\epsilon )$, for any given
rate $\epsilon >0$. In theoretical analysis it is standard to pay particular
attention to the support of the limit distribution $\mu (\epsilon )$ when $%
\epsilon $ approaches zero. We will refer to sets in the support of the
limit distribution as \emph{stochastically stable states}.

A set of states $Z^{\prime }\subseteq Z$ is \emph{absorbing} if once the ($%
\epsilon =0$) deterministic dynamic enters the set it will not leave it and
if it is minimal in the sense that it has no proper subset satisfying this
property. We are interested in absorbing sets in which play settles down to
a stationary distribution. Let $P_{zz^{\prime }}$ denote the probability of
transition from state $z$ to state $z^{\prime }$. Let $A$ be an absorbing
set of the model without noise. The \emph{basin of attraction} of $A$,
denoted by $D(A)$, is the set of all states from which the unperturbed
Markov process converges to a state in $A$. The characterization of the
long-run predictions of the stochastic model will rely on the calculation of
two useful concepts capturing the relative persistence over time of various
absorbing sets: the \emph{radius} and the \emph{coradius} of their
respective basins of attraction (see Ellison, 2000). While the formalization
of these concepts requires the use of some additional mathematical notation
(see SI Appendix), they are intuitive to grasp. Suppose the system is in an
absorbing set $A$. The \emph{radius} of the basin of attraction of $A$
corresponds to the minimum number of mutations necessary to leave the basin
of attraction. Next, compute the minimum number of mutations needed to reach
the basin of attraction of $A$, starting from an absorbing set outside $A$.
Do the same for all other absorbing sets outside $A$, and determine the
maximum of these numbers. This number is the \emph{coradius} of the basin of
attraction of $A$. The smaller the coradius, the likelier is the event that
simultaneous mutations shift the system from any absorbing state to some
state in $D(A)$. Ellison (2000) derived a sufficient condition for an
absorbing set to be uniquely selected by the stochastic evolutionary process
as the mutation rate vanishes: if the radius of the basin of attraction of
an absorbing set $A$ is larger than its coradius, all stochastically stable
sets are contained in $A$.

Intuitively, the radius provides a bound on the persistence of a set, while
the coradius provides a bound on its attractiveness. When $R(A)>CR(A)$ all
stochastically stable outcomes are in $A$. We will make use of the following
concept that is related to the coradius. The \emph{modified coradius, }$%
CR^{\ast }(A)$ captures the insight that, under certain conditions, large
evolutionary changes might occur more rapidly via a sequence of gradual
steps through a number of \textquotedblleft transient" steady states; see
Ellison (2000). The modified coradius is most useful in models with a large
number of deterministic steady states. It is computed by subtracting from
the coradius a correction term which depends on the number of intermediate
steady states along the evolutionary path and on the sizes of their basins
of attraction. When $\epsilon >0$, the invariant distribution $\mu (\epsilon
)$ exists and is globally stable (see SI Appendix). In addition, for (almost
all) histories of sufficient length, the weight a state receives in the
invariant distribution corresponds the the fraction of time the system
spends in this state (ergodicity). More formally, define the limit
distribution by $\mu ^{\ast }=\lim_{\epsilon \rightarrow 0}\mu (\epsilon )$.
The \emph{stochastically stable set} is defined by $Z^{\ast }=\{z\in Z:\mu
^{\ast }(z)>0\}$. Define $\mu ^{\ast }(A)=\Sigma _{z\in A}\mu ^{\ast }(z)$
with $\mu ^{\ast }(Z^{\ast })=1$. For any absorbing set\emph{\ }$A$, if $%
R(A)>CR^{\ast }(A)$, then $\mu ^{\ast }(A)=1$.

In short, the introduction of noise (mutations) allows us to make sharp
predictions about the long-run behavior of a dynamical system whose
deterministic version involves multiple steady states. Furthermore, as long
as the dynamical system satisfies the weak monotonicity (Darwinian)
property, its details regarding speed of adjustment do not matter. Its
long-run behavior is determined by the stochastic mutations. These allow for
multiple transitions across different steady states. However, in the
long-run, the steady state equilibrium with the largest basin of attraction
is played \textquotedblleft most of the time." The independence of the
theoretical predictions on the details of the payoff matrix and those of the
dynamical system involved is a desirable property since these factors are
unlikely to remain constant over evolutionary time horizons.

An example picture of the dynamics is shown in Figure 3. The states $0$ and $%
n=12$ refer to the two pure strategy Nash equilibria corresponding to
mycorrhizal fungi mutualisms (\textbf{AM}) and complete abandonment (\textbf{%
NM}). The three intermediary states in blue correspond to (mixed) Nash
equilibria, reflecting different levels of partial mycorrhizal fungi
mutualisms (\textbf{AMNM}).

\begin{figure}[!htpb]
\includegraphics[width=1.0\textwidth]{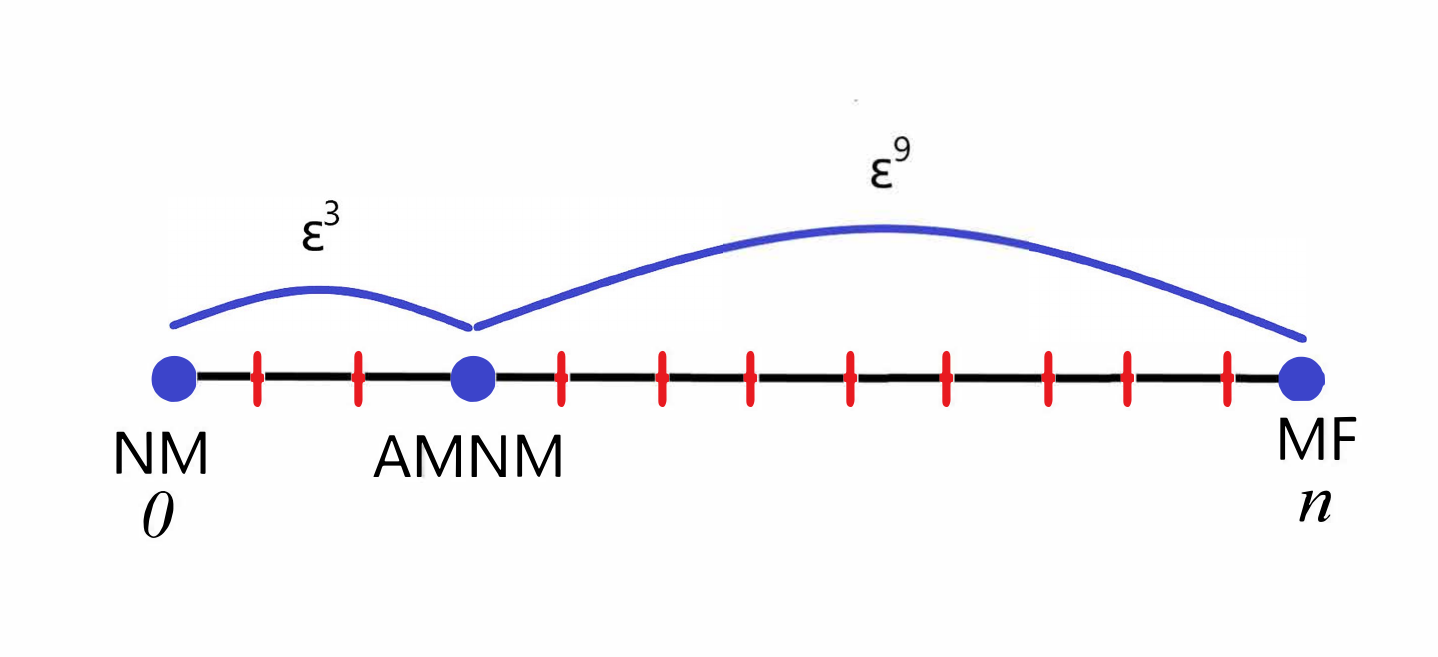}
 \centering
\caption{Coordination game: stochastic dynamics}
\label{fig:fig3}
\end{figure}

It is worth pointing out some qualitative features of the implied dynamics using the illustrative
example in Figure 3. First, all five circled states in Figure 3,
corresponding to the two pure (\textbf{AM} and \textbf{NM}) and the three
mixed (\textbf{AMNM}) Nash equilibria in this example, are steady states of
the dynamical system $z_{t+1}=b(z_{t})$ discussed earlier. Second, when
random mutations are introduced, the resulting Markov chain implies a
positive probability of transition from any state in $Z=\{0,1,...,12\}$ to
any other. As states that do not correspond to deterministic steady states
are transient, in determining the support of the invariant distribution it
is sufficient to consider the steady states, associated with the pure and
mixed Nash equilibria. In the above illustrative example, it takes a minimum
of seven mutations to escape the basin of attraction of the \textbf{AM}
steady state, an event that takes an expected time of $\frac{1}{\epsilon ^{7}%
}$. Similarly, escaping the basin of attraction of the \textbf{NM} steady
state requires a minimum of three mutations, an event that takes an expected
time of $\frac{1}{\epsilon ^{3}}$. Finally, as the basin of attraction of
either of the mixed steady states is a singleton, escaping either of them
requires a single mutation, an event that takes an expected time of $\frac{1%
}{\epsilon }$. When the mutation rate, $\epsilon $, is small, these expected
times have sharp predictions about the fraction of time that the system
spends at each state over a long enough history. More precisely, in the
above example the set of absorbing states is $Z^{\prime }=\{0,3,4,5,12\}$,
where only states $z=0$ and $z=12$ have a non-trivial basin of attraction.
We can then compute that $R(12)=7>4=CR^{\ast }(12)$, while $%
R(0)=3<8=CR^{\ast }(0)$. Thus, under sufficiently low mutation rates, state $%
z=12$, corresponding to the \textbf{AM} Nash equilibrium, would be observed
most frequently over a large time horizon. The use of modified co-radius
tightens the bounds by capturing the fact that the \textbf{AM} state is more
likely to be reached through intermediary \textbf{AMNM} states, a prediction
that is consistent with the evolutionary record.

It is also worth reiterating that these predictions follow directly from the
coordination structure of the underlying game and they do not depend on the
specific details of the dynamical system or the payoff matrix, which would
be very hard to infer in most applications. Of course, relative payoffs are
relevant, as they determine the relative size of the basins of attraction of
the various absorbing sets. Thus, the model is consistent with the
observation that a reduction in the benefit-to-cost ratio of the mutualism
contributes to the probability of mutualism abandonment. The biological
explanation is that this is because a reduced benefit-to-cost ratio of the
symbiosis should result in stronger natural selection to limit root
colonization. The mathematical explanation provided by our model is that a
reduction in the mutualism benefit-to-cost ratio reduces the size of the
basin of attraction of the \textbf{AM} absorbing set, thus requiring a
smaller number of mutations for the system to escape to an \textbf{AMNM}\
state.

\section{Quantitative explorations}

We will illustrate some of our main findings through a few representative
simulations of the basic model using the replicator dynamic, $B(z)$ in (\ref%
{R}). In principle, the model can be calibrated using phylogeny data to
obtain information about the respective basins of attraction of the three
types of Nash equilibria. For example, Maherali et al, 2016 calculated 4.23
transitions per million years from the \textbf{AMNM} to the \textbf{AM}
state per (see Figure 1). This amounts to approximately one such transition
per 236,407 years or, if we set a period to equal 10 years, to approximately
one transition per 23,600 periods. With a slight abuse in notation we now
let $z\in \lbrack 0,1]\ $stand for the fraction of players playing $s_{1}$
(the two pure Nash equilibria are then given by $z=0$ and $z=1$,
respectively). Our theoretical model implies that the expected time before
such a transition is $\frac{1}{\epsilon ^{z^{\ast }n}}$ periods. Given $%
\epsilon $ and $n$ the equation $\frac{1}{\epsilon ^{z^{\ast }n}}=23,600$
can thus be solved for the basin of attraction of the \textbf{AM} Nash
equilibrium, $z^{\ast }n=\ln (\frac{1}{23,600}-\epsilon )$. Similarly 0.054
transitions from the \textbf{AMNM} to the \textbf{NM} state per million
years (see Figure 1), imply (assuming again that a period equals 10 years)
that the basin of attraction of the \textbf{NM} Nash equilibrium is given by 
$(1-z_{\ast })n=\ln (\frac{1}{1,851,852}-\epsilon )$. Given the horizons
involved, such transitions are, of course, far too infrequent. We will thus
simulate a stylized numerical example in what follows in order to illustrate
some features of the model.

We divide the state space into three regions. For the simulations we need to
pick the number of players, $n$, the mutation rate, $\epsilon $, and the
time horizon, $T$, as well as the initial condition, $z_{0}\in \lbrack 0,1]$%
. We chose $n=50$, $\epsilon =.2$, and $T=600$. In addition, we chose the
indifference (mixed strategy) region to be between $z_{\ast }=0.2$ and $%
z^{\ast }=0.7$. The initial condition $z_{0}$ was chosen randomly according
to a uniform distribution across the (discretized) state space. The precise
payoffs used in the simulations are given in the SI Appendix. Recall that it
is only the difference in payoff sign that matters for the long-run model
predictions. The graphs in Figure 4 present representative results from four
different simulations. In all four graphs the horizontal axis measures time $%
t=1,...,600$, while the vertical axis represents the fraction of players
choosing strategy $s_{1}$ (\textbf{AM}). Thus, $z_{t}=1$ represents the 
\textbf{AM} equilibrium, $z_{t}=0$ represents the \textbf{NM} equilibrium,
while the region between the two horizontal lines, $z_{t}\in \lbrack
0.2,0.7] $, represents the mixed strategy Nash equilibrium points (\textbf{%
AMNM}). The coordination structure of the game implies that the basin of
attraction of the \textbf{AM} equilibrium is given by the interval $[0.7,1]$%
, while the basin of attraction of the \textbf{NM} equilibrium is the
interval $[0,0.2]$. This makes the \textbf{AM} equilibrium the one selected
to be played \textquotedblleft most of the time" over a sufficiently long
time horizon.

The first graph (top left) in Figure 4 represents a simulation where the
system stays in the basin of attraction of the \textbf{NM} Nash equilibrium.
Ongoing \textit{iid} mutations perturb the system away from $z=0$, but they
do not arise in sufficient magnitude over the simulation horizon to create
the market thickness that is necessary for the system to establish a
(partial or full) mutualism. Similarly, the second graph (top right)
represents a simulation where the system stays in the basin of attraction of
the \textbf{AM} Nash equilibrium. Again, ongoing \textit{iid} mutations
perturb the system away from $z=1$, but they are not sufficient for the
system to escape the basin of attraction of the symbiotic Nash outcome, and
the mutualism remains more or less intact for the duration.

More interestingly, the simulation in the third graph (bottom left) involves
a transition away from the basin of attraction of the \textbf{NM}
equilibrium, leading to an eventual emergence of the symbiotic \textbf{AM}\
Nash outcome. Consistent with the historical record in Maherali et al
(2016), the transition occurs when a sufficient number of mutations move the
system out of the basin of attraction of the \textbf{NM} Nash equilibrium
and in the mixed \textbf{AMNM} region, where the system spends some time
(6.3\% of the simulation time horizon) before it enters the basin of
attraction of the \textbf{AM} Nash equilibrium, where it stays for the
duration of the simulation. Of course, we know from our theoretical results
that several transitions would be observed over a sufficiently long time
horizon. However, these results also imply that the system will spend the
majority of time around the \textbf{AM} Nash equilibrium.

The mutations constantly perturb the system away from the Nash outcomes, but
most of the time they are too small in magnitude to sufficiently
increase/reduce market thickness, and evolutionary forces return the system
towards the corresponding Nash equilibrium. Occasionally, a large number of
mutations lead to a significant reduction/increase in market thickness, and
a corresponding reduction/increase in the fitness of the symbiotic strategy.
This is sufficient for the system to escape the basin of attraction of the
corresponding Nash outcome. The simulation in the fourth graph (bottom
right) involves the system starting in the basin of attraction of the 
\textbf{AM}\ equilibrium, then entering the basin of attraction of the 
\textbf{NM}\ Nash outcome through the mixed \textbf{AMNM} region, and
finally switching back to the \textbf{AM} outcome for the remaining duration
of the simulation. Again, consistent with the historical record, these
transitions occur through intermediary \textbf{AMNM }states, with the system
spending some time (8.2\% of the simulation horizon) in the intermediary 
\textbf{AMNM }mixed strategy Nash region. Of course, should the simulation
horizon increase, several (infrequent) transitions across the Nash
equilibria will be observed, leading to the long-run weights of the
respective Nash equilibria prescribed by the ergodic distribution $\mu $.

\begin{figure}[!htpb]
\includegraphics[width=1.0\textwidth]{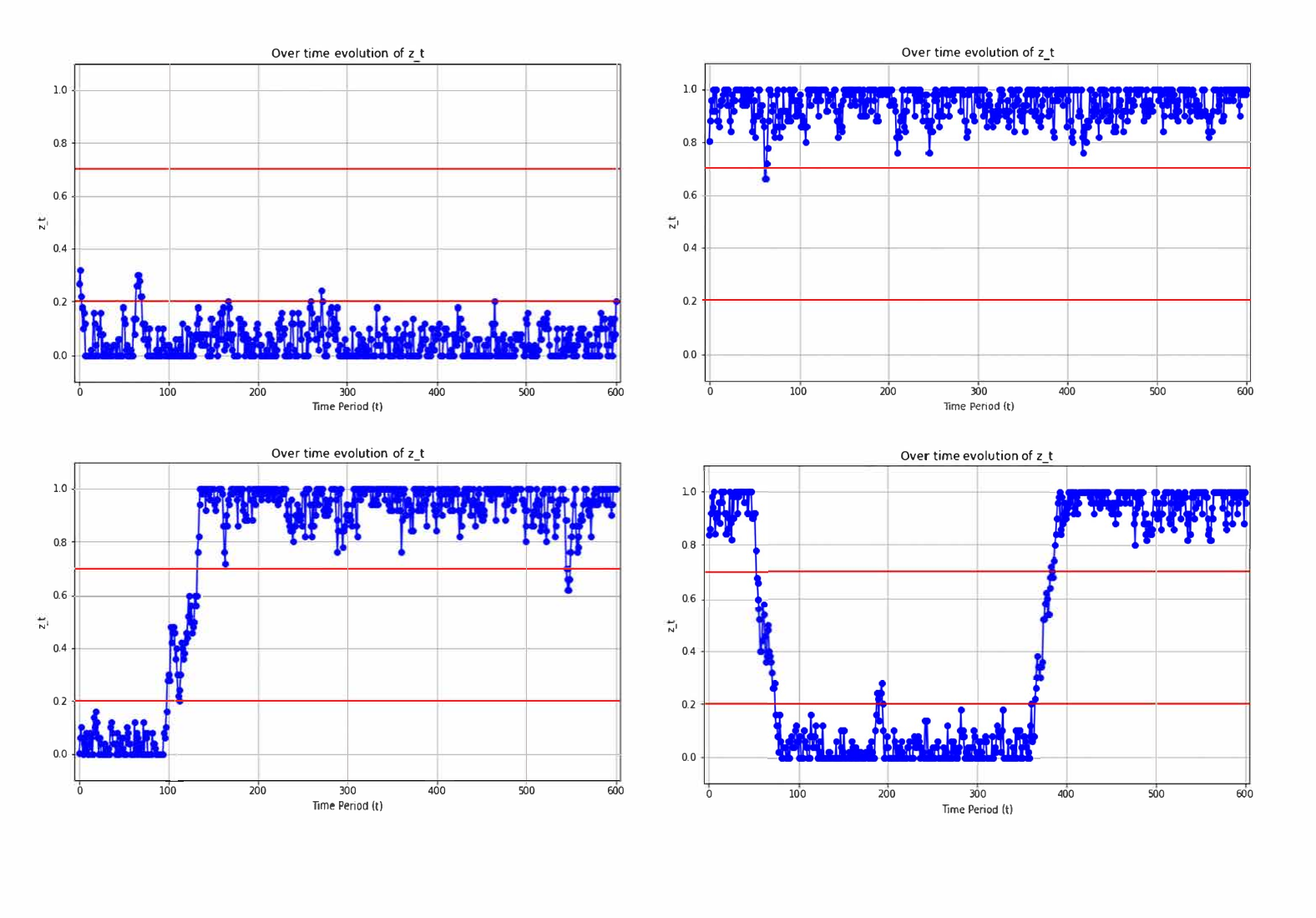}
 \centering
\caption{Coordination game simulated play}
\label{fig:fig4}
\end{figure}

Maherali et al (2016)
have pointed out that environmental and other changes that reduce the
benefit-to-cost ratio associated with the symbiosis increase the likelihood
of its abandonment. This insight is confirmed in the context of the
coordination game we study here since the change in the game's payoffs
resulting from a reduction in the benefit-to-cost ratio would lead to a
smaller basin of attraction of the \textbf{AM} Nash equilibrium and a
correspondingly larger basin of attraction of the \textbf{NM} Nash
equilibrium. As an example, below we represent the results from
representative simulations under the same conditions as before, but with the
basin of attraction of the \textbf{AM} equilibrium now given by the smaller
interval $[0.8,1]$, while the basin of attraction of the \textbf{NM}
equilibrium is the larger interval $[0,0.4]$.

\begin{figure}[!htpb]
\includegraphics[width=1.0\textwidth]{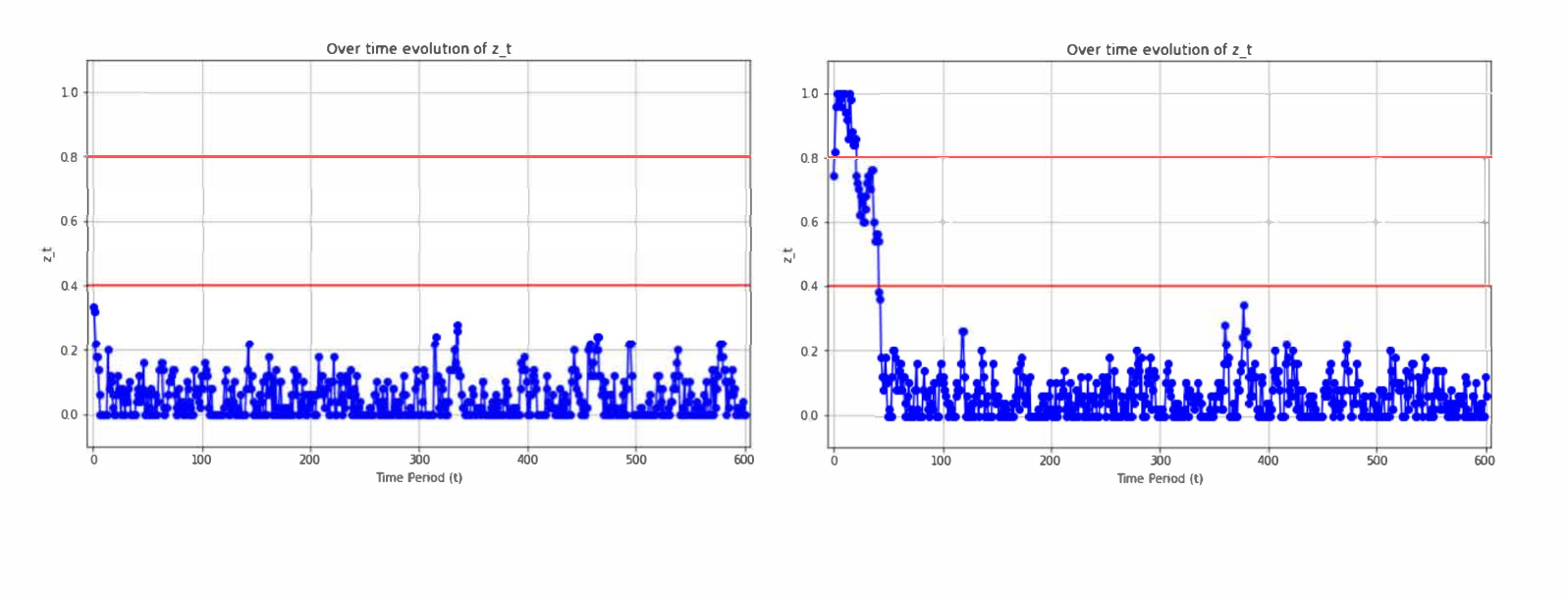}
 \centering
\caption{Lower symbiosis benefit/cost ratio}
\label{fig:fig5}
\end{figure}

Regardless of the initial condition, over a long time horizon play in this
case concentrates around the \textbf{NM} Nash equilibrium (Figure 5, left).
Once again, play moves repeatedly along the different basins of attraction
over evolutionary time. If the initial state is within the basin of
attraction of the \textbf{AM} Nash equilibrium, play will concentrate there
for a number of periods. However, as the basin of attraction of this
equilibrium shrinks, it becomes easier to escape, leading to a transition
(usually indirectly through the \textbf{AMNM} steady states) towards the 
\textbf{NM} equilibrium, where the system will spend a long time before it
eventually escapes again. Of course, the reverse conclusion would be reached
if the benefit-to-cost ratio associated with the symbiosis was to increase.
In that case, an increase in the size of the basin of attraction of the 
\textbf{AM} Nash equilibrium would make it the preferred outcome over a long
time horizon.

\section{Conclusions}

We employed techniques from large deviation theory to study stochastic
evolutionary dynamics in the context of a biological market. The
coordination game structure allowed us to build a rigorous theory of
mutualism emergence and abandonment that emphasizes the benefits of the
biological market size and does not rely on the possibility of
\textquotedblleft cheating" among participants. We illustrated our results
in the context of the seed plant-mycorrhizal fungi mutualism. Our model
captured some of the main findings in the formal ancestral reconstruction in
seed plants in Maherali et al (2016). Notably, our findings are consistent
with the observation that the mutualism has been abandoned and
re-established several times through evolutionary time, but it is more
likely to persist than be abandoned. We found that, while the symbiosis
emergence and abandonment could occur via direct transitions between the 
\textbf{AM} and the \textbf{NM} states, they are far more likely to occur
via an intermediate reversion through mixed \textbf{AMNM} states. Over a
long enough time horizon, the system spends most of the time in the Nash
equilibrium with the largest basin of attraction (\textbf{AM}). However, the
system will eventually escape and reach another absorbing set before it
escapes again over long evolutionary time horizons. Since exploitation of
plants by mycorrhizal fungi appears to be infrequent (Maherali et al. 2016),
we treated the fungi as passive in our analysis and concentrated on the
interactions among host plants, as in Halloway et al (2022). A more general
analysis would involve both sides of this biological market playing an
asymmetric coordination game. The application of the modified co-radius
formalizes the notion that large evolutionary changes are more likely when
they can be achieved by passing through a number of \textquotedblleft
transient" steady states. In this context, it explains why the \textbf{AM}
state is reached more frequently through the intermediary state (\textbf{AMNM%
}). A consequence of the multiple Nash equilibrium coordination game
framework is that the \textbf{AM}-plant mutualism might be (partially or
fully) abandoned by the plants even if it is overall superior for plant
fitness. Thus, coordination games giving rise to multiple, Pareto-ranked,
Nash equilibria might provide a rationale for why evolution can get stuck in
\textquotedblleft local fitness maxima," while global maxima might coexist.
This can have applications in other biological contexts.

\bigskip

\noindent \textbf{CODE.} The code used in our simulations can be found at:

\smallskip

GitHub: \textit{https://github.com/hemitheo/Mutualisms}

\bigskip
    
\noindent \textbf{FUNDING INFORMATION.} No outside funding was used in this research.

\bigskip
    
\noindent \textbf{ACKNOWLEDGMENTS.} Corresponding author: Ted Loch-Temzelides. We thank participants at the 2025 Mathematics of Uncertain Systems conference in Rimini, Italy the 16th Conference on Optimal Control and Dynamic Games in Vienna, Austria, and the AI, Data, and Decision Sciences Department at LUISS University (Rome, Italy) for comments. Pierre Loch provided computational assistance.
    
\bigskip

\section{References}

\begin{description}
\item Archetti, M., Scheuring, I., Hoffman, M., Frederickson, M. E., Pierce,
N. E., \& Yu, D. W. (2011). Economic game theory for mutualism and
cooperation. \textit{Ecology letters}, 14(12), 1300-1312.

\item Baker, A. Reef corals bleach to survive change. \textit{Nature} 411,
765--766 (2001). https://doi.org/10.1038/35081151

\item Blume, Andreas, and Ted Temzelides. On the geography of conventions. 
\textit{Economic Theory} 22: 863-873 (2003)

\item Bronstein, Judith L. The exploitation of mutualisms. \textit{Ecology
letters} 4.3 (2001)

\item Bronstein, Judith L., Ruben Alarc\'{o}n, and Monica Geber. The
evolution of plant--insect mutualisms. \textit{New Phytologist} 172.3:
412-428 (2006)

\item Boulotte, N., Dalton, S., Carroll, A. et al. Exploring the
Symbiodinium rare biosphere provides evidence for symbiont switching in
reef-building corals. \textit{ISME J} 10, 2693--2701 (2016).
https://doi.org/10.1038/ismej.2016.54

\item Culley, T. M., S. G. Weller, and A. K. Sakai. The evolution of wind
pollination in angiosperms. \textit{Trends in Ecology and Evolution}
17:361--369 (2002)

\item Dembo, Amir, and Ofer Zeitouni. \textit{Large deviations techniques
and applications}. Stochastic Modelling and Applied Probability Vol. 38.
Springer Science \& Business Media (2009)

\item Ellison, G.: Basins of attraction, long run equilibria, and the speed
of step--by--step evolution. \textit{Review of Economic Studies}\textsl{\ }%
67 (1)\textbf{, }17-45 (2000)

\item Ferriere, R\'{e}gis, et al. Cheating and the evolutionary stability of
mutualisms. \textit{Proceedings of the Royal Society of London. Series B:
Biological Sciences} 269.1493: 773-780 (2002)

\item Freidlin M. I., Wentzell A. D.: \textit{Random perturbations of
dynamical systems}. 1st Edition, New York: Springer Verlag (1984)

\item Friedman, J. Gone with the wind: understanding evolutionary
transitions between wind and animal pollination in the angiosperms. \textit{%
New Phytologist} 191:911--913 (2011)

\item Fudenberg D., Levine D.K.: \textit{The theory of learning in games}.
Cambridge, MA: MIT Press (1998)

\item Halloway, Abdel H., Katy D. Heath, and Gordon G. McNickle. When does
mutualism offer a competitive advantage? A game-theoretic analysis of
host--host competition in mutualism. \textit{AoB Plants} 14.2 (2022)

\item Jones, Emily I., et al. Cheaters must prosper: reconciling theoretical
and empirical perspectives on cheating in mutualism. \textit{Ecology letters 
}18.11: 1270-1284 (2015)

\item Josef Hofbauer and Karl Sigmund, \textit{Evolutionary Games and
Population Dynamics}, Cambridge University Press (1998)

\item Kandori, M., Mailath, G., Rob, R.: Learning, mutation, and long run
equilibria in games. \textit{Econometrica} 61 29-56 (1993)

\item Kauffman, S. and Levin, S.. Towards a general theory of adaptive walks
on rugged landscapes. \textit{Journal of theoretical Biology}, 128(1),
pp.11-45 (1987)

\item Levin, J., Miller, J. Broadband neural encoding in the cricket cereal
sensory system enhanced by stochastic resonance. \textit{Nature} 380,
165--168 (1996)

\item Maherali, Hafiz, Brad Oberle, Peter F. Stevens, William K. Cornwell,
and Daniel J. McGlinn. Mutualism persistence and abandonment during the
evolution of the mycorrhizal symbiosis. \textit{The American Naturalist}
188, no. 5: E113-E125 (2016)

\item McNickle, Gordon G., and Ray Dybzinski. Game theory and plant ecology. 
\textit{Ecology letters} 16, no. 4: 545-555 (2013)

\item Nash, J. F. Non-cooperative games: \textit{The Annals of Mathematics},
v. 54: 286-295 (1951)

\item No\"{e}, R., \& Hammerstein, P. (1994). Biological markets: supply and
demand determine the effect of partner choice in cooperation, mutualism and
mating. \textit{Behavioral ecology and sociobiology}, 35, 1-11

\item No\"{e}, R, and P. Hammerstein. Biological markets, \textit{Trends in
Ecology \& Evolution}, Volume 10, Issue 8, Pages 336-339,
https://doi.org/10.1016/S0169-5347(00)89123-5 (1995)

\item John Maynard Smith, \textit{Evolution and the Theory of Games},
Cambridge University Press (1982)

\item Palmer, Todd M., Maureen L. Stanton, Truman P. Young, Jacob R. Goheen,
Robert M. Pringle, and Richard Karban. \textquotedblleft Breakdown of an
Ant-Plant Mutualism Follows the Loss of Large Herbivores from an African
Savanna.\textquotedblright\ \textit{Science} 319, no. 5860 (2008): 192--95.
http://www.jstor.org/stable/20051975

\item Palmer, Todd M., Maureen L. Stanton, Truman P. Young, John S. Lemboi,
Jacob R. Goheen, and Robert M. Pringle. A Role for Indirect Facilitation in
Maintaining Diversity in a Guild of African Acacia Ants.\ \textit{Ecology}
94, no. 7: 1531--39 (2013). http://www.jstor.org/stable/23596942

\item Taylor P.D. and L.B. Jonker, Evolutionarily Stable Strategies and Game
Dynamics, \textit{Mathematical Biosciences}, 40(1--2), 145--156 (1978)

\item Vasconcelos, V\'{\i}tor V, Constantino, Sara M, Dannenberg, Astrid,
Lumkowsky, Marcel, Weber, Elke, Levin, Simon: Segregation and clustering of
preferences erode socially beneficial coordination. \textit{Proceedings of
the National Academy of Sciences} 118(50) (2021)

\item Werner, G. D., W. K. Cornwell, J. I. Sprent, J. Kattge, and E. T.
Kiers. A single evolutionary innovation drives the deep evolution

\item of symbiotic N2-fixation in angiosperms. \textit{Nature Communications}
5:4087 (2014)

\item Werner, G. D., J. H. C. Cornelissen, W. K. Cornwell, N. A.
Soudzilovskaia, J. Kattge, S. A. West, and E. T. Kiers, Symbiont switching
and alternative resource acquisition strategies drive mutualism breakdown, 
\textit{Proceedings of the National Academy of Sciences}, vol. 115 no. 20,
5229--5234 (2018)

\item Xu, Li, Denis Patterson, Ann Carla Staver, Simon Asher Levin, and Jin
Wang. Unifying deterministic and stochastic ecological dynamics via a
landscape-flux approach. \textit{Proceedings of the National Academy of
Sciences} 118, no. 24 (2021): e2103779118

\item Xu, L., Patterson, D., Levin, S.A. and Wang, J. Non-equilibrium
early-warning signals for critical transitions in ecological systems. 
\textit{Proceedings of the National Academy of Sciences}, 120(5),
p.e2218663120 (2023)

\item Young, P.~H.: The evolution of conventions. \textit{Econometrica} 61,
57-84 (1993)
\end{description}

\bigskip

\section{Supplemental information}

\subsection{Stochastic evolutionary dynamics}

Introducing noise or mutations turns the deterministic dynamical system, $%
z_{t+1}=b(z_{t})$, into a stochastic dynamical system: 
\begin{equation}
z_{t+1}=b(z_{t})+z_{1t}-z_{2t}
\end{equation}%
where $z_{1t}$, $z_{2t}$ follow binomial distributions: $z_{1t}\sim
B(n-b(z_{t}),\epsilon /2)$, and $z_{2t}\sim B(b(z)_{t},\epsilon /2)$. The
stochastic dynamical system gives rise to a Markov chain with transition
matrix:

\begin{equation}
P=[z_{t+1}=j|z_{t}=i]
\end{equation}
where $i$, $j=0,...,n$. Let $\mu (\epsilon )$ be an \textit{invariant
distribution} associated with $P$; i.e., $\mu (\epsilon )P=\mu (\epsilon )$.
Since $\epsilon >0$, every element of $P$ is strictly positive. As is well
known, this is a sufficient condition for the existence and uniqueness of $%
\mu (\epsilon )$. In addition, $\mu $ satisfies the following properties:

\textit{Global stability: }$\forall \mu _{0}$, $\lim_{t\rightarrow \infty
}\mu _{0}P^{t}\rightarrow \mu $

\bigskip

\textit{Ergodicity:} Define:

\begin{equation*}
I_{i}(z_{t}):=\left\{ 
\begin{array}{ccc}
1, & \text{if} & z_{t}=i \\ 
0, & \text{if} & z_{t}\neq i%
\end{array}%
\right.
\end{equation*}

Then, $\forall z_{0}$, $\lim_{T\rightarrow \infty }\frac{1}{T}\sum
{}_{t=1}^{T}I_{i}(z_{t})\rightarrow \mu _{i}$ almost surely.

Define the limit distribution by $\mu ^{\ast }=\lim_{\epsilon \rightarrow
0}\mu (\epsilon )$. The \emph{stochastically stable set} is defined by $%
Z^{\ast }=\{z\in Z:\mu ^{\ast }(z)>0\}$. Let $\mu ^{\ast }(A)=\Sigma _{z\in
A}\mu ^{\ast }(z)$ with $\mu ^{\ast }(Z^{\ast })=1$. We next discuss
characterizing the invariant distribution. First, we have the following
notion for long-run outcomes of the deterministic ($\epsilon =0$) dynamic, $%
b(z)$.

\noindent \textbf{Definition 3: }\textit{A set of states} $Z^{\prime
}\subseteq Z$ \textit{is absorbing if (i) for all} $z^{\prime }\in Z^{\prime
}$, $z\notin Z^{\prime }$, $P_{z^{\prime }z}=0$, \textit{and (ii)} $
{\nexists}Z^{\prime \prime }\subset Z^{\prime }$, $Z^{\prime \prime }\neq
Z^{\prime }$ \textit{s.t. (i) holds for} $Z^{\prime \prime }$.

The first condition requires that once the process enters the absorbing set,
it will not leave it. The second condition requires that absorbing sets are
minimal. We are interested in absorbing sets in which play settles down to a
stationary distribution. Let $P_{zz^{\prime }}$ denote the probability of
transition from state $z$ to state $z^{\prime }$. Let $A$ be an absorbing
set of the model without noise. The \emph{basin of attraction} of $A$,
denoted by $D(A)$, is the set of all states from which the unperturbed
Markov process converges to a state in $A$ with probability one, 
\begin{equation}
D(A)=\{z\in Z|\Pr (\exists \tau ^{\prime }\,\text{\textit{s.t.} }\,z_{\tau
}\in A\;\forall \tau >\tau ^{\prime }|z_{0}=z)=1\}
\end{equation}%
For any set $A$, the \emph{radius} of $D(A)$, is the number of mutations
necessary to leave the set, starting from a state in $A$ (see Ellison,
2000). Let $c(z,z^{\prime })$ be the number of mutations needed for the
system to transit from state $z$ to state $z^{\prime }$. That is, $c(\cdot )$
measures the \emph{transition cost} between these states. Define a \emph{path%
} by a finite sequence $(z_{1},\ldots ,z_{k})$ of distinct states. The \emph{%
cost} of such a path is defined by 
\begin{equation}
c(z_{1},\ldots ,z_{k})=\Sigma _{\tau =1}^{k-1}c(z_{\tau },z_{\tau +1})
\end{equation}%
Formally, the \emph{radius} of $A$ is the least costly path leading from any
state in $A$ to some state outside the basin of attraction of $A$.

\noindent \textbf{Definition 4}: \textit{The radius of the basin of
attraction of a collection of absorbing sets }$A$ \textit{is} 
\begin{equation}
R(A)=\min_{(z_{1},\ldots ,z_{k})}c(z_{1},z_{2},\ldots ,z_{k})\text{ s.t. }%
z_{1}\in A\text{, }z_{k}\notin D(A)
\end{equation}%
The path $(z_{1},\ldots ,z_{k})$ defining the radius of $D(A)$ describes the
\textquotedblleft cheapest" way out of that set. Formally, the \emph{coradius%
} of the basin of attraction of a collection of absorbing sets is defined by
the number of mutations necessary to reach this set from the state where the
minimum number of mutations required to reach $D(A)$ is maximized.

\noindent \textbf{Definition 5}: \textit{The coradius of the basin of
attraction of a set of absorbing sets }$A$ \textit{is}: $CR(A)=\max_{z_{1}%
\notin A}\min_{(z_{1},\ldots ,z_{k})}c(z_{1},\ldots ,z_{k})$\textit{\ such
that} $z_{k}\in D(A)$.

When $R(A)>CR(A)$ all stochastically stable sets are in $A$ and that the
expected waiting time until a limit set $A$ is reached is at most $%
O(\epsilon ^{-CR(A)})$ (see Ellison, 2000).

\noindent \textbf{Definition 6}: \textit{The modified coradius of the basin
of attraction of a collection of absorbing sets }$A$\textit{\ is given by:} 
\begin{equation}
CR^{\ast }(A)=\max_{z^{1}\notin A}\min_{(z^{1},\ldots
,z^{k})}\{c(z_{1},\ldots ,z_{k})-\sum_{l=2}^{L-1}R(Z_{l})\}\text{ \textit{%
s.t. }}z_{k}\in D(A)
\end{equation}%
\textit{where} $\{Z_{l}\}$ \textit{is the sequence of absorbing sets through
which }$(z_{1},\ldots ,z_{k})$ \textit{passes.}

A tighter bound on the expected waiting time until a limit set $A$ is
reached is then at most $O(\epsilon ^{-CR^{\ast }(A)})$. A sufficient
condition for a set of states to be stochastically stable is that the radius
of its basin of attraction exceeds the modified coradius. We restrict
attention to such states. When $R(A)>CR^{\ast }(A)$ all stochastically
stable sets are in $A$ and that the expected waiting time until a limit set $%
A$ is reached is at most $O(\epsilon ^{-CR^{\ast }(A)})$ (see Ellison, 2000).

\bigskip

\subsection{Sketch of the algorithm steps used in simulations}

\noindent Fix number of players: $n$

\noindent Fix mutation rate: $\epsilon $

\noindent Fix number of periods: $t=0,...,T$

\noindent Fix initial condition: $z_{0}=$ random fraction in $[0,1]$

\noindent Define fraction of $s_{1}$ players at $t$: $z_{t}\in \lbrack 0,1]$

\noindent Define fraction of $s_{2}$ players at $t$: $1-z_{t}$

\noindent Define payoff from playing $s_{1}$ at $t$: $u(s_{1},z_{t})$

\noindent Define payoff from playing $s_{2}$ at $t$: $u(s_{2},z_{t})$

\noindent Set $u_{t}(z_{t}):=(s_{1},z_{t})-u(s_{2},z_{t})$ where, for all $t$%
,

$u(z_{t}):=\left\{ 
\begin{array}{ccc}
-1, & \text{if} & z_{t}<z_{\ast } \\ 
0, & \text{if} & z_{t}\in \lbrack z_{\ast },z_{\ast }] \\ 
+1, & \text{if} & z_{t}>z^{\ast }%
\end{array}%
\right. $

\noindent For all $t$, set: $u(s_{1},z_{t}):=\left\{ 
\begin{array}{ccc}
2, & \text{if} & z_{t}<z_{\ast } \\ 
2.5, & \text{if} & z_{t}\in \lbrack z_{\ast },z_{\ast }] \\ 
3, & \text{if} & z_{t}>z^{\ast }%
\end{array}%
\right. ,u(s_{2},z_{t}):=\left\{ 
\begin{array}{ccc}
3, & \text{if} & z_{t}<z_{\ast } \\ 
2.5, & \text{if} & z_{t}\in \lbrack z_{\ast },z_{\ast }] \\ 
2, & \text{if} & z_{t}>z^{\ast }%
\end{array}%
\right. $

\begin{equation}
z_{t+1}=z_{t}\left[ \frac{u(s_{1},z_{t})}{%
z_{t}u(s_{1},z_{t})+(n-z_{t})u(s_{2},z_{t})}\right] +\epsilon
(1-z_{t})-\epsilon z_{t}
\end{equation}

\bigskip

\noindent Starting with $z_{0}$, use the iterative process above to compute $%
z_{t}$ for every integer time period $t$, $t=0,...,T$

\noindent Compute the fraction of time the system spends in each region $%
R_{1}$, $R_{2}$, $R_{3}$

\noindent Compute the fraction of transitions from $z=1$ to $z=0$ that
occurred directly, versus through $z\in \lbrack z_{\ast },z^{\ast }]$

\end{document}